\documentclass[11pt,leqno]{article}

\usepackage{xspace}
\usepackage{comment}
\usepackage{amsmath}
\usepackage{amsfonts}

\makeatletter%
\def\nottoobig#1{{\hbox{$\left#1\vcenter to1.111\ht\strutbox{}\right.\n@space$}}}
\makeatother%

\makeatletter%

\newcount\hour  \newcount\minutes  \hour=\time  \divide\hour by 60
\minutes=\hour  \multiply\minutes by -60  \advance\minutes by \time
\def\mmmddyyyy{\ifcase\month\or Jan\or Feb\or Mar\or Apr\or May\or Jun\or Jul\or
  Aug\or Sep\or Oct\or Nov\or Dec\fi \space\number\day, \number\year}
\def\hhmm{\ifnum\hour<10 0\fi\number\hour :%
  \ifnum\minutes<10 0\fi\number\minutes}
\def\Draft{{\it Draft of \mmmddyyyy}}

\def\ps@jtsheadings{%
\def\@oddhead{\it\rightmark\hfil\rm\thepage}%
\def\@oddfoot{\hfil\Draft}%
\if@twoside%
\def\@evenhead{\rm\thepage\hfil\it\leftmark}%
\def\@evenfoot{\Draft\hfil}%
\else
\let\@evenhead\@oddhead%
\let\@evenfoot\@oddfoot%
\fi%
}
\def\ps@jtsplain{%
\def\@oddhead{\hfil\Draft}%
\def\@oddfoot{\hfil\rm\thepage\hfil}%
\let\@evenfoot\@oddfoot%
\if@twoside \def\@evenhead{\Draft\hfil} \else \let\@evenhead\@oddhead \fi
}

\def\chaptermark#1{\markboth{\thechapter.\ #1}{\thechapter.\ #1}}%
\def\sectionmark#1{\markright{\thesection.\ #1}}

\def\section{\@startsection {section}{1}{\z@}
    {3.5ex plus1ex minus.2ex}{2.3ex plus.2ex}{\Large\bf}}
\def\subsection{\@startsection{subsection}{2}{\z@}
    {3.25ex plus1ex minus.2ex}{1.5ex plus.2ex}{\large\bf}}
\def\subsubsection{\@startsection{subsubsection}{3}{\z@}
    {3.25ex plus1ex minus.2ex}{1.5ex plus.2ex}{\normalsize\bf}}
\def\paragraph{\@startsection{paragraph}{4}{\z@}
    {3.25ex plus1ex minus.2ex}{1em}{\normalsize\bf}}
\def\subparagraph{\@startsection{subparagraph}{4}{\parindent}
    {3.25ex plus1ex minus.2ex}{1em}{\normalsize\bf}}
\makeatother%

\makeatletter \@beginparpenalty=10000 \makeatother

\def\underl#1 {\leavevmode\let\first=\relax\underli #1 }
\def\underli#1 {\ifx&#1\let\next=\relax\unskip
                \else\let\next=\underli\first\ulinebox{#1}\fi\let\first=\undersp\next}
\def\undersp{\penalty50\ulinebox{\space}\penalty50}
\def\ulinebox#1{\vtop{\hbox{\strut#1}\hrule}}%
\def\unice#1 {\underl #1 & }

\def\desclabel#1{\bf #1\hfil}
\def\desc{\list{}{%
\labelwidth= \leftmargin
\advance \labelwidth by -\labelsep
\let \makelabel=\desclabel}}



\makeatletter %

\newcommand{\nat}{\mathbb{N}}
\newcommand{\mapping}{\rightarrow}

\def\union{\,\cup}
\def\inter{\,\cap}

\newlength{\filength}
\newsavebox{\gcbox}
\sbox{\gcbox}{\framebox[\filength]{\rule{0ex}{2ex}}}

\newlength{\leftjustindent}
\newlength{\@leftjustindent}
\def\leftjust{\let\\\@leftjustcr\let\end\@endleftjust
  \addtolength{\@leftjustindent}{\leftjustindent} \vcenter\bgroup
\halign\bgroup \hbox to\displaywidth{
\rule{\@leftjustindent}{0ex}$\displaystyle##$\hfill }\crcr }
\def\endleftjust{\crcr\egroup\egroup\endgroup}
\def\@endleftjust#1{\crcr\egroup\egroup\@checkend{#1}\endgroup}
\def\@leftjustcr{\crcr}

\newtheorem{theorem}{Theorem}[section]
\newtheorem{corollary}[theorem]{Corollary}
\newtheorem{claim}[theorem]{Claim}

\newcommand{\qedblob}{\mbox{\rule[-1.5pt]{5pt}{10.5pt}}}
\def\literalqed{{\ \nolinebreak\hfill\mbox{\qedblob\quad}}}

\def\qed{\literalqed}

\newtheorem{fact}[theorem]{Fact}

\newcommand{\singlespacing}{\let\CS=
\@currsize\renewcommand{\baselinestretch}{1}\tiny\CS}
\newcommand{\singlespacingplus}{\let\CS=
\@currsize\renewcommand{\baselinestretch}{1.25}\tiny\CS}
\newcommand{\doublespacing}{\let\CS=
\@currsize\renewcommand{\baselinestretch}{1.75}\tiny\CS}
\newcommand{\draftspacing}{\let\CS=
\@currsize\renewcommand{\baselinestretch}{2.0}\tiny\CS}
\newcommand{\normalspacing}{\singlespacing}
\makeatother%

\mathcode`\0="0030      %
\mathcode`\1="0031
\mathcode`\2="0032
\mathcode`\3="0033
\mathcode`\4="0034
\mathcode`\5="0035
\mathcode`\6="0036
\mathcode`\7="0037
\mathcode`\8="0038
\mathcode`\9="0039
\singlespacingplus

\newtheorem{definition}[theorem]{Definition}

\makeatletter
\clubpenalty=\@highpenalty
\widowpenalty=\@highpenalty
\makeatother

\makeatletter
\newcommand{\niceonespacing}{\let\CS=\@currsize\renewcommand{\baselinestretch}{1.1}\tiny\CS}\newcommand{\nicetwospacing}{\let\CS=\@currsize\renewcommand{\baselinestretch}{1.2}\tiny\CS}
\newcommand{\nicethreespacing}{\let\CS=\@currsize\renewcommand{\baselinestretch}{1.3}\tiny\CS}
\newcommand{\singlespacingplusplus}{\let\CS=\@currsize\renewcommand{\baselinestretch}{1.35}\tiny\CS}
\newcommand{\nicefourspacing}{\let\CS=\@currsize\renewcommand{\baselinestretch}{1.4}\tiny\CS}
\newcommand{\nicefivespacing}{\let\CS=\@currsize\renewcommand{\baselinestretch}{1.5}\tiny\CS}
\newcommand{\nicesixspacing}{\let\CS=\@currsize\renewcommand{\baselinestretch}{1.6}\tiny\CS}
\makeatother

\makeatletter
\def\@cite#1#2{[#1\if@tempswa , #2\fi]}
\makeatother

\makeatletter
\def\@citex[#1]#2{\if@filesw\immediate\write\@auxout{\string\citation{#2}}\fi
  \def\@citea{}\@cite{\@for\@citeb:=#2\do
    {\@citea\def\@citea{,\linebreak[0]}\@ifundefined
       {b@\@citeb}{{\bf ?}\@warning
       {Citation `\@citeb' on page \thepage \space undefined}}%
\hbox{\csname b@\@citeb\endcsname}}}{#1}}
\makeatother

\makeatletter
\def\ps@thesis{\def\@oddhead{\hfil\rm\thepage\hfil}\def\@oddfoot{}\def\@evenhead{\hfil\rm\thepage\hfil}\def\@evenfoot{}\def\chaptermark##1{}\def\sectionmark##1{}}
\makeatother

\newcommand{\p}{{\rm P}}

\newcommand{\np}{{\rm NP}}
\newcommand{\qp}{{\rm QP}}

\newcommand{\bpp}{{\rm BPP}}

\normalspacing

\newcommand{\sigmastar}{\ensuremath{\Sigma^\ast}}
\newcommand{\pisnp}{\ensuremath{\p=\np}}

\newcommand{\calo}{\ensuremath{{\cal O}}}

\newcommand{\condition}{\,\nottoobig{|}\:}

\newenvironment{block}{\begin{list}{\hbox{}}{\leftmargin 1em
    \itemindent -1em \topsep 0pt \itemsep 0pt \partopsep 0pt}}{\end{list}}

\normalspacing
\title{Almost-Everywhere Superiority \\ for Quantum 
Polynomial Time\protect\footnote{\protect\singlespacing{}Supported
in part by grant
NSF-INT-9815095/\protect\linebreak[0]DAAD-315-PPP-g\"u-ab.}}

\author{
Edith Hemaspaandra\footnote{\protect\singlespacing{}Dept.~of 
Comp.~Sci.~,~Rochester Institute of
Technology, Rochester,
NY 14623.  \mbox{eh@cs.rit.edu}.}
\and 
Lane A. Hemaspaandra\footnote{\protect\singlespacing{}Department of 
Computer Science,
University of Rochester, Rochester,
NY 14627.
\mbox{lane@cs.rochester.edu}.}
\and 
Marius Zimand\footnote{\protect\singlespacing{}Department of Computer 
and Information Sciences,
Towson University,
Towson, MD 21252 and Department of Computer Science, University of Bucharest, Bucharest, Romania.
\mbox{mzimand@saber.towson.edu}.}
} %

\date{April 28, 2000}

\makeatletter
\def\@listI{\leftmargin\leftmargini \parsep 4.5pt plus 1pt minus 1pt\topsep
6pt plus 2pt minus 2pt \itemsep  2pt plus 2pt minus 1pt}

\let\@listi\@listI
\@listi
\makeatother

\makeatletter %
 \newcommand{\setoffdisplay}{\rule{5.9in}{1pt}}

\makeatother



\newcommand{\prob}{{\rm Prob}}

\begin{document}

\typeout{WARNING:  BADNESS used to supress reporting.  Beware.}
\hbadness=3000%
\vbadness=10000 %

\bibliographystyle{alpha}

\pagestyle{plain}

{\singlespacing\maketitle}

\begin{abstract}
  
  Simon~\cite{sim:j:power-of-quantum} as extended by Brassard and
  H{\o}yer~\cite{bra-hoy:c:exact-simon} shows that there are tasks on
  which polynomial-time quantum machines are exponentially faster than
  each classical machine infinitely often.  The present paper shows
  that there are tasks on which polynomial-time quantum machines are
  exponentially faster than each classical machine almost everywhere.

\end{abstract}

\normalspacing

\sloppy

\section{Introduction}\label{s:intro}
One issue of broad importance in the area of
quantum computing is to gain an understanding of exactly what
potential quantum computers hold, i.e., what superiority over
classical computers they offer.
Work of Simon~\cite{sim:j:power-of-quantum}, 
as extended by Brassard and 
H{\o}yer~\cite{bra-hoy:c:exact-simon},\footnote{An alternative
algorithm to that of 
Brassard and H{\o}yer was obtained by Mihara and 
Sung~\protect\cite{mih-sun:c:qp-for-simon} and by 
Beals~\protect\cite{hoy:perscomm:991024-bra-hoy-alternate-algorithms}.}
is often cited as
evidence for the potential 
superiority 
of quantum computation 
over classical computation.  
Their work shows that for
computation relative to a black-box function (also sometimes
referred to as a promise function)
there are problems on
which exact (i.e., worst-case polynomial time with
zero error probability, see, e.g.,~\cite{bra-hoy:c:exact-simon})
polynomial-time quantum computation
is infinitely often
exponentially faster than each 
deterministic---or even bounded-error
probabilistic---classical computer solving the problem.

Berthiaume and
Brassard~\cite{ber-bra:j:quantum-oracles} raised and studied
the issue of whether one can obtain far more decisive separations:
separations where quantum computation is superior on all but a 
finite number of inputs.   That is, they sought to bring to 
the quantum-versus-classical computation question the very strong
type of separation known in complexity theory 
as almost-everywhere 
separations~\cite{ges-huy-sel:cOUTbyges-huy-sei-REALLYsei:ae,ges-huy-sei:j:ae-hierarchies,all-bei-her-hom:j:nondet-ae-hier}.
Berthiaume and
Brassard~\cite{ber-bra:j:quantum-oracles} 
obtained the remarkable result that
there are
  tasks that can be done in exact exponential time on quantum 
  machines but on which each classical deterministic machine requires
  double exponential time almost everywhere.

However, neither 
the Berthiaume-Brassard result nor the technique of 
its proof works 
for quantum polynomial time.  
(Such results tend to be far
harder to obtain for small time classes than for large ones.)
Also, their result
is for
deterministic (not bounded-error probabilistic) classical computing. 
Additionally, the Simon result leaves open the possibility 
(which in fact is the case that holds, as first discussed by 
Berthiaume and Brassard) that 
for some classical computers solving the problem there are 
infinitely many
inputs
on which
quantum computing is {\em not\/} interestingly faster on
these problems. {\em In fact,
quantum computing in Simon's construction is 
superior of classical computing on only
an exponentially small portion of the inputs.}

In contrast, the present paper
shows that there are problems on which exact quantum polynomial-time 
computing is exponentially superior to classical computing {\em almost
  everywhere}.  In particular, {\em we show that 
for computation
relative to a black-box function 
there are problems solved in
exact polynomial-time by quantum computers but on which every
deterministic---or even bounded-error probabilistic---classical 
computer solving the problem requires exponential time on all but a
finite number of inputs.}

\section{History and Discussion}

This section provides 
a more detailed history and discussion 
of the background and related results than 
does Section~\ref{s:intro}.  Reading this 
section is not needed to understand the results 
of Section~\ref{s:results-ae}.

\subsection{Simon and Infinitely-Often
  Superiority for Quantum Computing}

Tremendously exciting new models of computation---quantum computing
and DNA-computing---have become one strong
focus of theoretical computer science research.  Researchers dearly
want to know whether these models, at least in certain settings, offer
computational properties (most particularly, quick run-time)
superior to what is offered by classical computational models.

Of course, even such an exciting model as quantum computing has 
limitations (see, e.g., the elegant lower-bound approach of Beals
et al.~\cite{bea-buh-cle-mos-dew:c:lower}).  
However, let us here consider the
highlights of what is known suggesting the superiority of quantum
computing.  The three most famous lines of work are those of 
Shor~\cite{sho:j:quantum-factoring},
Grover~(\cite{gro:c:quantum-database}, see
also~\cite{boy-bra-hoy-tap:j:tight-quantum-searching}),
and Simon~(\cite{sim:j:power-of-quantum}, see 
also~\cite{bra-hoy:c:exact-simon}).

Grover shows that quantum computing can do certain search problems at a
quadratically faster exponential speed than one intuitively would
expect in classical computing.  Shor shows that factoring (and other
interesting problems) can be done in expected polynomial time in the
quantum model.  These are both 
undeniably impressive results.  However, note that
it is at least plausible that classical, deterministic computing can
seemingly have the effect of searching 
through huge numbers of possibilities very quickly 
(for example, testing satisfiability) and can
factor very quickly; for example, if $\pisnp$, NP-like search 
problems\footnote{\protect\singlespacing{}To be fair to Grover,
his result can plausibly be viewed instead as a black-box
result.  
The key issue is whether the predicate, $C(S)$, that he uses
should be viewed as some polynomial-time evaluation 
or as a black-box 
predicate.
He does not have to address this issue (his 
motivating example, SAT, satisfies the former but
a
parenthetical remark in his paper 
suggests the latter), as 
his results are valid either way and as 
he is 
improving the upper bound rather than establishing any lower bounds.
In any case, note that 
in contrast to Simon's and the present paper's
exponential superiority results, Grover's 
algorithm beats the obvious brute-force deterministic 
algorithm by a quadratic factor.}
and factoring are easily in P
(though clearly not by going through all the 
possibilities anywhere near a brute-force way).

In contrast, Simon~\cite{sim:j:power-of-quantum}
shows that for computing with respect to a
black-box function
there are problems for which quantum polynomial-time bounded-error
computing provably {\em is\/} exponentially faster than classical
deterministic computing or even classical bounded-error computing.
Brassard and H{\o}yer~\cite{bra-hoy:c:exact-simon}
improved the upper bound to obtain that for
computing with respect to a black-box function there are problems for
which exact quantum 
polynomial-time computing is exponentially faster than classical
deterministic computing or even classical bounded-error computing; in
particular, there are problems in 
exact polynomial time 
(which we will
refer to as $\qp$, see~\cite{bra-hoy:c:exact-simon}, though it sometimes
is denoted EQP) 
such that each bounded-error
classical Turing machine solving them requires exponential
time on infinitely many inputs.

\subsection{Limitations of Simon's Result}

As described in the previous subsection, Simon-Brassard-H{\o}yer show,
for computing with respect to a black-box function, the
infinitely-often exponential superiority of exact quantum polynomial-time
computing over classical deterministic computing
(and even over classical bounded-error computing), on a particular
problem.  Since we will always speak of computing with respect to a
black-box function that may have a promise, 
we will henceforth stop mentioning that and take
it to be implicit from context as is standard in the literature.

Are there any worries or limitations to Simon's 
work?
Simon (when one
tightens his upper bound to QP via the work of
Brassard-H{\o}yer) gives an ``infinitely often'' result: a problem
that is in QP but such that each classical bounded-error machine
solving the problem takes
exponential time on infinitely many inputs.  However, ``infinitely
many'' says no more than it seems to.  In fact, for Simon's problem,
there are classical deterministic machines that solve the problem 
essentially 
instantly (i.e., in $n+1$ steps on inputs of length $n$) on the vast
majority of inputs---in fact, on all but one input of each length.

{\em So, even though Simon proves infinitely-often superiority, in fact for
his problem the superiority occurs only on an exponentially thin
portion of inputs.  In contrast, the present paper 
achieves exponential superiority for 
exact quantum polynomial time on {\em all\/} sufficiently long
inputs (so, for example, each classical machine for the problem will take
subexponential time on at most a finite set of inputs).}

This is well-motivated, as one issue of broad importance in the area of
quantum computing is to gain an understanding of exactly what
potential quantum computers hold, i.e., what superiority over
classical computers they offer.

Thus, it is not surprising that the relation between 
classical and quantum computing is currently 
under intense scientific scrutiny. 
We briefly mention some other works that have disclosed
various facets of this relation and that exhibit, 
in different settings or different time classes, 
superiority in favor of quantum computing.
As noted above, an
early paper of Berthiaume and
Brassard~\cite{ber-bra:j:quantum-oracles} raised the important issue
of almost-everywhere hardness for quantum computing, and showed that
there are tasks that can be done in exact exponential time on quantum
machines but on which each classical deterministic machine requires
double exponential time almost everywhere.  In contrast, 
our paper achieves almost-everywhere separation for exact quantum
{\em polynomial time}, and handles bounded-error as well
as deterministic classical machines.   
Ambainis and de~Wolf~\cite{dew:perscomm:991012-quantum-average-case} 
have informed us
that, independently of the work of this paper, which
first appeared 
in~\cite{hem-hem-zim:tButWithExactDayToShowIndep:almost-everywhere-quantum},
they have studied average-case separations with 
respect to the uniform 
distribution, their theorem
related to this paper first appearing
in~\cite{amb-dew:tWithExactDateForIndepPoint:average-quantum}
(the earlier ``Version 1'' of that report 
does not contain the related result;
see also~\cite{amb-dew:c:average-quantum}).    
What is the relationship between their work and ours?
Of course,
almost-everywhere separation implies average-case
separation in the standard sense, and thus our main result certainly
implies average-case separation with 
respect to the uniform distribution.  However, their paper is 
formally incomparable to ours as the models are exceedingly different
(some ways in favor of the strength of their results, and some ways
in favor of the strength of our results), for example
(in their section related
to this paper, their Section~4):
(1)~their fast quantum algorithms are Las Vegas-type algorithms (and
thus some computation paths may take far longer than polynomial time)
rather
than exact quantum algorithms,
(2)~their input is exponentially long relative to their ``$n$''
and so they are actually
distinguishing
quantum {\em logarithmic query complexity\/} from 
classical {\em polynomial query complexity},
(3)~we are computing a total 
(for the specific oracle obtained in our proof) 
non-Boolean function and they are computing a
total Boolean function (note that 
due to work of Beals et al.~\cite{bea-buh-cle-mos-dew:c:lower}
it is known that in the query complexity model, which is 
the model of 
Beals et al.~and of Ambainis and de~Wolf but not of the present paper,
superpolynomial {\em query complexity\/} 
gaps between quantum and classical
computation cannot ever be obtained 
for total Boolean functions;  but keep in
mind that this does not speak directly to the issue of time
complexity gaps in standard, non-(random access)-type-time-counting models),
(4)~their model of input and queries is different than 
ours as in some sense their input is their oracle 
(and so uniform distribution must be viewed
in this context) and their
notion (see 
also~\cite{bea-buh-cle-mos-dew:c:lower}) 
of query complexity essentially measures accessing the input
itself, and 
(5)~they study average-case complexity but we study almost-everywhere
separations. 
Finally, we mention that in the important (but completely different)
area of communication complexity,
Raz~\cite{raz:c:quantum-communication-complexity} has shown that for
promise problems there is an exponential gap between quantum
communication complexity (which in particular is logarithmic on his 
problem) and classical probabilistic communication complexity 
(which he gives a lower-bound on as a root of the input size).

\section{Almost-Everywhere Superiority for Quantum Polynomial Time}\label{s:results-ae}

Let us start by explicitly stating where we will go.
Recall that, as is common, we will throughout this
paper denote quantum exact polynomial time
(see~\cite{bra-hoy:c:exact-simon})
by $\qp$,
though it sometimes
in earlier papers is 
denoted EQP\@.
Recall that what Simon's main theorem states (again, using here
the Brassard-H{\o}yer improvement of the upper bound
to QP) is the following.

\begin{theorem}\label{t:sbh} (\cite[Theorem~3.4]{sim:j:power-of-quantum}
augmented by~\cite{bra-hoy:c:exact-simon})
\quad
There is a constant $\epsilon > 0$ and a (function) oracle
$\calo$ relative to which there is a language $B$ in 
exact quantum polynomial time 
such that each bounded-error classical Turing machine
accepting $B$ requires time more than $2^{\epsilon n}$ 
on infinitely many inputs.
\end{theorem}

What we will prove is the following result, which extends the
superiority from merely infinitely often to instead almost everywhere.

\begin{theorem}\label{t:our-main}
There is a constant $\epsilon > 0$ and a (function) oracle
$\calo$ relative to which there is a problem $B$ computable in exact quantum polynomial time
such that each bounded-error classical Turing machine
computing $B$ requires time more than $2^{\epsilon n}$ 
on all but a finite number of inputs.
\end{theorem}

It follows immediately that this problem also demonstrates the
almost-everywhere superiority of quantum computation over
deterministic computation, when computing relative to a black-box
function.

\begin{corollary}\label{c:determ}
There is a constant $\epsilon > 0$ and a (function) oracle
$\calo$ relative to which there is a problem $B$ computable in exact quantum polynomial time
such that each deterministic classical Turing machine
computing $B$ requires time more than $2^{\epsilon n}$ 
on all but a finite number of inputs.
\end{corollary}

Some comments are in order regarding Theorem~\ref{t:our-main}.  First,
we should mention that the computational task on which we prove almost
everywhere exponential superiority for quantum computing is, in
contrast with Simon's task, a function rather than a language.
Second, we should explicitly define what we mean by a probabilistic 
function.

\begin{definition}\label{def:bounded-error-function}
We say a function $f$ is bounded-error Turing computable
in time $T(n)$ (i.e., is in ${\rm BPTIME[}T(n){\rm{}]}\,$) iff there is 
an $\epsilon > 0$ and
a probabilistic Turing machine $M$ running in time $T(n)$ such
that, on each $x\in\sigmastar$, $$\prob ( M(x) = f(x) )  \geq
1/2 + \epsilon.$$
If $M$ is a probabilistic Turing machine satisfying the above relation, we say
that $M$ has {\em error probability} at most $1/2 - \epsilon$.
\end{definition}

Finally, we review
a bit about Simon's result, as his result motivated
our work,
as we should credit him for 
the connections between his construction and ours, and
as it is important to point out 
why the obvious transformation of his result does {\em not}
give the result we seek.

The key construction used by Simon is described in the statement of
the following result.

\begin{theorem}\label{t:simon-underlying} 
(\cite[Theorem~3.3]{sim:j:power-of-quantum})\quad
Let $\calo$ be a (function) oracle constructed as follows:
for each $n$, a random $n$-bit string $s(n)$ and a random bit $b(n)$
are uniformly chosen from $\{0,1\}^n$ and $\{0,1\}$,
respectively.  If $b(n) = 0$, then the function $f_n:
\{0,1\}^n \rightarrow \{0,1\}^n$ chosen 
for $\calo$ to compute on $n$-bit queries is a random function
uniformly distributed over permutations on $\{0,1\}^n$;
otherwise it is a random function uniformly distributed over
two-to-one functions such that $f_n(x) = f_n(x \oplus s(n))$ for all
$x$, where $\oplus$ denotes bitwise exclusive-or.
Then any PTM (probabilistic Turing machine) that queries $\calo$
no more than $2^{n/4}$ times cannot correctly
guess $b(n)$ with probability greater than $(1/2) + 2^{-n/2}$,
over choices in the construction of 
$\calo$.\footnote{\protect\singlespacing{}\label{f:simon}The statement
here is taken
exactly from Simon.  There are some 
informalities in Simon's statement---the 
fact that what independence is assumed 
is not explicitly stated and 
that the case ``$b(n) = 1 \wedge s(n)=0^n$'' won't give an (exactly-2)-to-1
function.
}
\end{theorem}

Simon's ``test language'' that, based on this oracle, gives one the
lower-bound of Theorem~\ref{t:sbh} is quite simply the issue of
testing the bit described above, that is, the test language that is in
QP but on which bounded-error $2^{\epsilon n}$-time
classical Turing machines
all err on infinitely many inputs is $\{1^n \condition b(n) = 1\}$.

It might be very tempting to exactly adopt the oracle $\calo$ of
Simon, but using instead of his test language the new test
language: $\widehat{L} = \{w \condition b(|w|)=1\}$.  This change
attempts to ``smear'' the difficulty of $1^n$ onto all strings of
length $n$, and even attempts to achieve the language analog of 
our desired result.

Unfortunately, this provably does not work.  Why?  A PTM can use the
information in the input to (very rarely, but often enough) help it
guess $s(n)$, in particular, certainly when it holds that
both $b(n) = 1$ and the input happens to be
 $s(n)$.\footnote{\protect\singlespacing{}Just to be 
explicit here for 
absolute clarity, and assuming in light of the comments
in Footnote~\ref{f:simon} that we never allow the choice
``$b(n) = 1 \wedge s(n) = 0^n$,'' consider the PTM that on each input $w$ does:

{\samepage
\noindent
\{$n=|w|$;\\
$a =$ output of oracle $\calo$ on input $0^n$; \\
$b =$ output of oracle $\calo$ on input $0^n \oplus w$;\\
if $a=b$ and $w \not \in 0^*$
then output ``$b(|w|)=1$ and $s(|w|) = w$'' else output ``$b(|w|)=0$.''\}
} %

This machine will, on an infinite number of inputs $w$ (on
each length $n$ for which $b(n) = 1$, on the input that equals $s(n)$;
and for each length $n$ for which $b(n) = 0$, on all inputs),
correctly determine
$b(|w|)$ with probability one (relative to the choices of the PTM).
Of course, this machine is not correctly accepting $\widehat{L} = \{w
\condition b(|w|)=1\}$, but the machine is enough to show that keeping
Simon's oracle $\calo$ and just adopting the test set $\widehat{L}$ does
not establish Simon's Theorem~\ref{t:simon-underlying}
in the analogous case that applies here, i.e., where any length $n$
string $w$ may be the input.
We note that the PTM given does a bit more than this;  on each $n$ with
$b(n) = 1$, we have at least one input on which the PTM not only
knows $b(n)$ but even discovers $s(n)$.
} %

So, our construction takes a different tack.  Intuitively speaking,
the above problem should be removed if we increase the information
content of the xor-bitmask well beyond that which input strings can
give away.  To achieve this, we double the information content 
of the xor-bitmask string, and demand that our functions discover 
this string.

\bigskip

\noindent {\bf Proof of Theorem~\ref{t:our-main}:}\quad 
We consider function oracles $A$ of the following form: $A$ is a collection  of functions
$(f_{n,A})_{n \in \nat^{+}}$  with the following properties:
\begin{itemize}
\item[(i)] $f_{n,A} : \{0, 1\}^n \mapping \{0, 1\}^{n-1}$,
\item[(ii)] $f_{n,A}$ is 2-to-1,
\item[(iii)] there is a string $s_{n,A}$ in $\{0,1\}^n - \{0^n\}$ such that for all $x$ of length $n$, 
$f_{n,A}(x \oplus s_{n,A}) = f_{n,A}(x)$.
\end{itemize}

Let ${\cal A}$ be the set of all such oracles. One can easily induce a probability measure on ${\cal A}$. Indeed, ${\cal A}$ is the 
product of the sets $({\cal A}_i)_{i \in \nat^{+}}$, where, for each $i \in \nat^{+}$, ${\cal A}_i$ is the set of all functions
$f$ mapping $\{0, 1\}^i$ into $\{0, 1\}^{i-1}$ and having the properties (i), (ii), and (iii). 
On each set ${\cal A}_i$ we consider the probability measure given by the uniform distribution 
and then we consider the product measure 
on ${\cal A}$. This is identical 
to choosing,  for each $n$ independently, $f_{n,A}$ according to the uniform distribution over all functions with the
properties (i), (ii), and (iii). All the probabilistic considerations that follow will be relative to this probability 
measure. It is important to observe that choosing $f_{n,A}$ uniformly at random amounts to the selection of a random
string $s$ of length $n$ and to the independent selection of a random permutation from $\{0, 1\}^{n-1}$ to $\{0, 1\}^{n-1}$ 
that dictates how the $2^{n-1}$ pairs $(u, u \oplus s)_{u \in \{0, 1\}^n}$, ordered in some canonical way and identified
with $\{0, 1\}^{n-1}$, are mapped into $\{0, 1\}^{n-1}$. 

Let $A \in {\cal A}$. We define $g_A$, a function mapping strings of length $n$ into strings of length $2n$, by
\[
g_A(w) = s_{2|w|,A},
\]
$\mbox{i.e.}$, $g_A(w)$ is the unique string $s$ with the property that for all $x$ of  length $2|w|$,
\[
f_{2|w|,A} ( x \oplus s) = f_{2|w|,A} ( x ).
\]

It follows from the work of Brassard and H{\o}yer~\cite{bra-hoy:c:exact-simon} that there is a machine 
running in quantum polynomial time that computes $g_A$ for all $A \in {\cal A}$.

Later in this proof, we will prove the following claim.

\begin{claim}
\label{cl:determ}
There is a set of oracles ${\cal B}_0$ having measure one in ${\cal A}$,
such that for every $A \in {\cal B}_0$ and every deterministic oracle 
machine $M$ the following holds: for almost every input $w$,
$M^A$ either runs for more than $2^{|w|/4} - 2$ steps or does not
calculate $g_A(w)$. 
\end{claim}

To move to bounded error probability machines, 
we invoke the techniques that Bennett and Gill~\cite{ben-gil:j:prob1} 
used to prove $\p^A = \bpp^A$ relative to a random oracle. An 
adaptation of their method shows the following.

\begin{claim}
\label{cl:ben-gil}
There is
a set of oracles ${\cal B}_1$ having measure one in ${\cal A}$, 
such that for any probabilistic oracle machine $N$ and for any $A$
in ${\cal B}_1$,  there exists a deterministic oracle machine $M$
with the following property: if $N^A$
computes a function $h$ with bounded error probability 
(in the sense of Definition~\ref{def:bounded-error-function}), then on 
all sufficiently long inputs $w$ on which $N^A$ runs in time $2^{|w|/5}$,
$M^A(w) = h(w)$ and $M^A$ runs in time $2^{|w|/4}-2$.
\end{claim}

Claims~\ref{cl:determ} and~\ref{cl:ben-gil} imply 
Theorem~\ref{t:our-main} (with 
${\cal O} \in {\cal B}_0 \inter {\cal B}_1$, $\epsilon = 1/5$,
and $B = g_{\cal O}$). 
For suppose for a contradiction that  there exists a
probabilistic oracle machine $N$ and ${\cal O} \in 
{\cal B}_0 \inter {\cal B}_1$,
such that $N^{\cal O}$ bounded-error
computes $g_{\cal O}$ in the sense of
Definition~\ref{def:bounded-error-function} and such that $N^{\cal O}$ 
runs in time $2^{|w|/5}$ for infinitely many inputs $w$.
Then, by Claim~\ref{cl:ben-gil}, there exists a deterministic
oracle machine $M$ that, for infinitely many inputs $w$,
calculates $g_{\cal O}(w)$ and runs in time $2^{|w|/4} - 2$.  
But that contradicts Claim~\ref{cl:determ}.

For completeness and since there are some differences between our context
and the one in the paper of Bennett and Gill~\cite{ben-gil:j:prob1},
we will prove Claim~\ref{cl:ben-gil} in detail. In the proof, we 
will assume that all the oracles $A$ are in ${\cal A}$.
If $N$ is a probabilistic oracle machine, $A \in {\cal A}$ 
an oracle, and $N^A$ computes a function $h$
with bounded error probability, we will write $N^A(w)$ to denote $h(w)$.

Let $N$ be a probabilistic oracle machine, let $A$ be 
an oracle (in ${\cal A}$), and let $r$ be a rational number
such that $0 \leq r < 1/2$ and
$N^A$ computes a function with error probability at most
$r$. Let us fix, as a parameter, a positive integer $k$.

If we iterate $N$ on input $w$ a polynomial number of times
(the polynomial depends on $k$ and $r$),
and, on each computation path, output the majority output
among the polynomially many computations of $N$ if a majority
output exists (if not, we (arbitrarily) output 0),
we get a new machine $N_{k,r}^{\prime A}$ that, on all oracles $A$ on which $N$ has
error probability at most $r$, computes the same function as 
$N^A$ but with probability error at most $(1/k) 2^{-(2|w|+1)}$
for every input $w$.

For all oracles $A$, $N_{k,r}^{\prime A}$
runs in time  $(1/2) 2^{|w|/4.5}$ on all sufficiently long inputs $w$
on which $N^A$ is
running in time $2^{|w|/5}$. Also note that $N_{k,r}^{\prime A}$
queries strings of length at most $2^{|w|/5}$ on all the inputs $w$
on which $N^A$ runs in time $2^{|w|/5}$.
From $N_{k,r}'$, we build a deterministic machine $M_{N, k, r}$ as follows.
Machine $M_{N, k,r}$ on input $w$ simulates $N_{k,r}'$
on input $w$ and each time $N'$ requires a random bit for doing a
probabilistic step, $M_{N, k, r}$ takes this bit to
be the first bit of $f_{t,A}(0^t)$, where  $t$ is
the smallest integer $> 2^{|w|/5}$ such that
$0^t$ has not been queried before during the simulation on input $w$.
It is easy to check that for all strings $w$ that are long enough,
if $N_{k,r}^{\prime A}$
on $w$ runs in time $(1/2) 2^{|w|/4.5}$, then $M_{N, k,r}^A$ on $w$ runs in time
$2^{|w|/4}- 2$.
For each $w$, and each rational $r$ with $0 \leq r < 1/2$,
let $E_{N, k, r, w}$ be the
class of oracles $A$ on which $N^A$ on input $w$ runs in $2^{|w|/5}$
steps and has error probability at most $r$, and on which
$M_{N, k,r}^A(w) \neq N_{k,r}^{\prime A}(w)$.
Let $U_1, \ldots, U_s$ be all the partial functions
defined on the 
strings of length at most $2^{|w|/5}$ such that for all
$i \in \{1, \ldots, s\}$, $N^{U_i}$ on input $w$
runs in $2^{|w|/5}$ steps with error probability at most $r$.
For an oracle $A$, let $A_{low}$ 
denote its restriction  to the strings of length at most $2^{|w|/5}$. Then
\[
\prob_{A} (A \in E_{N,k,r,w}) = \sum_{i=1}^{s} \prob_A(M_{N, k,r}^A(w)
\neq N_{k,r}^{\prime A}(w)~|~ A_{low} = U_i) \cdot \prob_A(A_{low} = U_i).
\]
Now, $\prob_A(M_{N,k,r}^A(w) \neq N_{k,r}^{\prime A}(w)~|~ A_{low} = U_i)$
is the probability that $M_{N, k,r}^A(w) \neq N_{k,r}^{\prime A} (w)$
given that the regular queries of both machines are answered according
to $U_i$. Since the only queries besides those stipulated by $U_i$ that
are involved
in the conditioned event ``$M_{N, k,r}^A(w) \neq N_{k,r}^{\prime A}$" are those
used by $M_{N, k,r}$ to simulate 
the random bits used by $N_{k,r}^{\prime U_i}$ on $w$,
it follows that the above conditioned probability is the
error probability of $N_{k,r}^{\prime U_i}(w)$ which is at most
$(1/k)2^{-(2|w|+1)}.$ It follows that $\prob_A(A \in E_{N,k,r,w}) 
\leq (1/k) 2^{-(2|w|+1)} \cdot \sum_{i=1}^{s} \prob_A(A_{low}=U_i)
\leq (1/k) 2^{-(2|w|+1)}.$

Let $E_{N,k,r}$ denote the set $\union_w E_{N,k,r, w}$,
where the union is taken over all
strings $w$. 
Note first that if $A \not \in E_{N,k,r}$ and if $N^A$ has probability error
at most $r$, then $M_{N, k,r}^A(w) = N^A(w)$
on all inputs $w$ on which $N^A$ runs in $2^{|w|/5}$ steps.
We have that $\prob_A(A \in E_{N,k,r}) \leq \sum_{w} \prob_A (A \in E_{N,k,r,w}) =
(1/k) \sum_w 2^{-(2|w| + 1)} = 1/k.$ Therefore the measure
of $\inter_{k \geq 1} E_{N,k,r}$ is zero and thus the measure of
${\cal A} - \inter_{k \geq 1} E_{N,k,r}$ is one.
We take ${\cal B_1} = \inter_{N,r} ( {\cal A} - \inter_{k \geq 1}
E_{N,k,r})$, where the first intersection is taken over all
probabilistic Turing machines $N$ and rationals $r$ such that $0 \leq r < 1/2$.
The set ${\cal B_1}$ has measure one because it is a numerable intersection
of sets of measure one. 

Let $N$ be a probabilistic Turing machine and $A$ be an oracle
in ${\cal B_1}$ such that
$N^A$ has bounded error probability at most $r$
for rational $r$ with $0 \leq r < 1/2$.
It follows that $A \in {\cal A} - E_{N,k,r}$ for some $k$. 
On all sufficiently long inputs $w$ on which $N^A$ runs in 
$2^{|w|/5}$ steps, $M_{N,k,r}$ runs in time
$2^{|w|/4} -2$ and $M_{N,k,r}^A(w) = N^A(w)$.
This completes the proof of Claim~\ref{cl:ben-gil}.

We now prove Claim~\ref{cl:determ}, that is, we have to
show that there is a 
set of oracles ${\cal B}_0$ having measure one in ${\cal A}$,
such that for every $A \in {\cal B}_0$ and every deterministic oracle 
machine $M$ the following holds: for almost every input $w$,
$M^A$ either runs for more than $2^{|w|/4} - 2$ steps or does not
calculate $g_A(w)$. 

Thus, let $M$ be a deterministic oracle machine that attempts
to calculate $g_A$. We modify $M$ so that 
at the end of its computation, having a string $s$ 
on its output tape, it asks the oracle $A$ for the values of $f_{|s|,A}(0^{|s|})$ and $f_{|s|,A}(s)$. 
Let $M'$ be the modified machine. The reason for this modification is so we are sure there is a ``collision'' 
if $M$ has the correct string $s$, as we will now make formal and clear. We say that for an oracle $A$, two strings $x$ and $y$ collide if 
$f_{|x|,A}(x) = f_{|y|,A}(y).$ Let us fix an input $w$ and let $n = |w|.$  Observe that
\begin{multline}
\label{queries}
\prob_A(M^A \mbox{ runs at most $2^{n/4}-2$ steps and calculates } g_A(w) ) \leq \\
\prob_A(M^{\prime A} \mbox{ queries at most $2^{n/4}$ strings and }  \\ 
\mbox{ two queried strings of length $2n$ collide with respect to $A$}),
\end{multline}
because if $M^A$ is correct on $w$, then $M^{\prime A}$ at the end of its computation will ask $0^{2n}$ and $s_{2n, A}$ 
and these will collide.

We assume without loss of generality that for each $z$ and for each oracle $A$ it holds
that $M^{\prime A}(z)$
does not query the same string twice during its run. Let $x_1, x_2, \ldots, x_k$ be, in the order in which they are
queried, the strings that $M'$ queries on input $w$. Of course, $k$ and the set of strings 
are random variables (in other words they depend on the oracle $A$). We will show the following fact.
\begin{fact}\label{f:main-fact}
$p_w =_{\mbox{\it def}} \prob_A (k \leq 2^{|w|/4}$ and there is a collision
for a pair of strings of length $2|w|$ in 
$\{x_1, \ldots, x_k\}) \leq 2^{-1.4|w|}$.
\end{fact}

Assuming that the fact holds, we have 
\begin{equation}
\begin{split}
\sum_{w \in \{0,1\}^*} p_w & =
\sum_{\ell=0}^{\infty} \sum_{w \in \{0, 1\}^\ell} p_w 
  = \sum_{\ell=0}^{\infty} 2^\ell \cdot 2^{-1.4\ell} 
  = \sum_{\ell=0}^{\infty} 2^{-0.4\ell} 
  < \infty.
\end{split}
\end{equation}
By the Borel-Cantelli Lemma and taking into account~(\ref{queries}) it follows that
\begin{multline*}
\prob_A(\mbox{for infinitely many inputs $w$, $M^A(w)$ makes at 
most $2^{|w|/4}-2$ steps} 
\\ 
\mbox{and computes } 
 g_A(w)) = 0.
\end{multline*}
Since there are a countable number of  deterministic oracle machines $M$, we obtain that
\begin{multline*}
\prob_A( \mbox{there exists } M \mbox{ that, on infinitely many inputs $w$, runs at most $2^{|w|/4}-2$ steps} \\
\mbox{and that computes }
 g_A(w) )= 0.
\end{multline*}
Consequently,
\begin{multline}
\label{e:prob1}
\prob_A( \mbox{for all } M, \mbox{ $M^A$, on almost every input $w$, either runs more than $2^{|w|/4}-2$ steps }\\
\mbox{ or does not compute }
g_A(w)) = 1,
\end{multline}
which is the desired assertion.

We still must prove  Fact~\ref{f:main-fact}.
In this proof, for brevity, collisions will always refer to strings of length $2n$ and will always
be with respect to the oracle $A$.
We will drop the subscript from the functions $f$, with the understanding that the missing subscript 
is equal to the length of the argument. We will also write $\prob(\ldots)$ for $\prob_A(\ldots)$ when this is clear 
from the context.

Decomposing the event ``$k \leq 2^{n/4}$ and collision in $\{x_1, \dots, x_k\}$" into mutually
disjoint events, we have
\begin{multline}
\label{eq1}
\prob(k \leq 2^{n/4} \mbox{ and collision in } \{x_1, \ldots, x_k \})  = \\
 \prob(k \leq 2^{n/4} \mbox{ and collision in } \{x_1, x_2\}) + \\
 \prob(k \leq 2^{n/4} \mbox{ and } x_3 \mbox{  collides with $x_1$ or $x_2$} \mbox{ and no collision in $\{x_1, x_2\}$}) + \\
 \ldots + \\
 \prob(k \leq 2^{n/4} \mbox{ and } x_k \mbox{ collides with $x_1$ or $x_2$ or $\ldots$ or $x_{k-1}$} \mbox{ and no collision 
in $\{x_1, \ldots, x_{k-1}\}$}) \\
\leq \sum_{j=2}^{2^{n/4}} \prob(x_j \mbox{ collides with $x_1$ or $x_2$ or $\ldots$ or $x_{j-1}$} \mbox{ and no collision 
in $\{x_1, \ldots, x_{j-1}\}$}),
\end{multline}  
with the convention that events involving some $x_j$ with $j > k$ are empty (and thus have probability zero).
We look at the general term in the above sum.
\begin{multline} \label{eqsplit}
\prob(x_j \mbox{ collides with $x_1$ or $x_2$ or $\ldots$ 
or $x_{j-1}$} \mbox{ and no collision 
in $\{x_1, \ldots, x_{j-1}\}$}) = \\
\sum \prob (x_j \mbox{ collides with $x_1$ or $\ldots$ }  \mbox{ or $x_{j-1}$} 
  \mbox{ and no collision 
in $\{x_1, \ldots, x_{j-1}\}$} ~| \\
(\forall i \in \{1, \ldots, j  \})[x_i = u_i]
\mbox{ and } (\forall i \in \{1, \ldots, j -1 \}) [f_A(u_i) = a_i])
\times \\
\prob( (\forall i \in \{1, \ldots, j  \})[x_i = u_i]\mbox{ and } (\forall i \in \{1, \ldots, j -1 \}) [f_A(u_i) = a_i]),
\end{multline}  
where the sum is taken over all $j$-tuples $(u_1, \ldots, u_j)$ of distinct strings in $\{0, 1\}^*$ 
(that we consider as potential queries of $M'$ on $w$) 
and over all possible answers $(a_1, \ldots, a_{j-1})$ to the queries $u_1, \ldots, u_{j-1}$ such that
the possible answers of length $2n-1$ are distinct (these are answers to queries of length $2n$ and 
they are distinct because
there is no collision in $\{u_1, \ldots, u_{j-1}\}$). Let us fix a tuple $(u_1, \ldots, u_j)$ of
possible distinct queries and a tuple $(a_1, \ldots, a_{j-1})$ of possible answers as above and 
let us consider the probability
\begin{multline*}
\prob (x_j \mbox{ collides with $x_1$ or $\ldots$ }  \mbox{ or $x_{j-1}$} 
 \mbox{ and no collision 
in $\{x_1, \ldots, x_{j-1}\}$} ~| \\
(\forall i \in \{1, \ldots, j  \})[x_i = u_i] \mbox{ and } (\forall i \in \{1, \ldots, j -1 \}) [f_A(u_i) = a_i]),
\end{multline*}
which is of course equal to
\begin{multline}
\label{eq2}
\prob (u_j \mbox{ collides with $u_1$ or $\ldots$ }  \mbox{ or $u_{j-1}$} 
\mbox{ and no collision 
in $\{u_1, \ldots, u_{j-1}\}$} ~| \\
(\forall i \in \{1, \ldots, j  \})[x_i = u_i ] \mbox{ and } (\forall i \in \{1, \ldots, j -1 \})[ f_A(u_i) = a_i]).
\end{multline}
Note that the condition ``no collision in $\{u_1, \ldots, u_{j-1}\}$" is subsumed
by the condition ``$(\forall i \in \{1, \ldots, j-1\})[f_A(u_i) = a_i]$" because the answers $a_i$,  
for $i=1, \ldots, j-1$, are distinct with respect to those of them of length $2n-1$. 
The conditions $f_A(u_i) = a_i$, for $i = 1, \ldots, j-1$, completely determine whether it is the case that for all 
$i \in \{1, \ldots, j\}$, the $i$-th query is $u_i$, $\mbox{i.e.},$ whether for all $i \in \{1, \ldots, j\}$, $x_i = u_i$.
 Thus the event $\{\mbox {no collision 
in $\{u_1, \ldots, u_{j-1}\}$}  \mbox{ and } (\forall i \in \{1, \ldots, j \}) [x_i = u_i] \mbox{ and } 
(\forall i \in \{1, \ldots, j-1 \}) [f_A(u_i) = a_i]\}$ is either empty or is equal to the event 
$\{(\forall i \in \{1, \ldots, j -1 \}) [f_A(u_i) = a_i] \}$.
If it is empty, the probability in equation~(\ref{eq2}) is zero (by the standard convention
regarding conditional probabilities). In the other case, the probability in equation~(\ref{eq2}) is equal to
\begin{multline*}
\prob(u_j \mbox{ collides with } \{u_1, \ldots, u_{j-1} \}~|~ 
(\forall i \in \{1, \ldots, j -1 \}) [f_A(u_i) = a_i]) \\
= \frac{ \prob(u_j \mbox{ collides with } \{u_1, \ldots, u_{j-1} \} \mbox{ and } (\forall i \in \{1, \ldots, j -1 \}) [f_A(u_i) = a_i])}
{\prob( (\forall i \in \{1, \ldots, j -1 \}) [f_A(u_i) = a_i]) }. 
\end{multline*}  
If $|u_j| \neq 2n$ the above conditional probability is zero. So, we will consider that $|u_j| = 2n$.
Let $U = \{u_i~|~i \in \{1, \ldots, j-1\} \mbox { and } |u_i| = |u_j| = 
2n \}$ and let $W = \{u_1, \ldots, u_{j-1}\} - U$\@.  Note 
that $||U||$, the cardinality of $U$, is at most $j-1$. Observe also that
$u_j$ cannot collide with elements from $W$ and that the events ``$u_j$ collides with some element in $U$
and $f_A(u_i) = a_i$,
for all $u_i$ in $U$ " and ``$f_A(u_i) = a_i$,
for all $u_i$ in $W$" are independent. The events ``$f_A(u_i) = a_i$,
for all $u_i$ in $U$" and ``$f_A(u_i) = a_i$,
for all $u_i$ in $W$" are also independent 
(the choices made in the construction of the oracle at different lengths are independent).
Therefore the probabilities involving strings $u \in W$ cancel and it remains to evaluate
\begin{equation}
\label{eq3}
\frac{ \prob(u_j \mbox{ collides with } \{u_i~|~ u_i \in U \} \mbox{ and } (\forall u_i \in U) [f_A(u_i) = a_i]) }
{\prob( (\forall u_i \in U) [f_A(u_i) = a_i] )}.
\end{equation}
The events in the above equation depend on the choices of the string $s$ and of the permutation
that determines $f_{2n, A}$, and these two choices are independent, as we have observed when we built the probability
measure. Let us focus on the event appearing in the numerator. For this event to hold, the string $s$, which is
responsible for the collisions, must be chosen so as to make $u_j$ collide 
with one of $\{u_i~|~u_i \in U\}$, 
and so as to prevent any collision in $U$ (because the ``answers" $a_i$  to the ``queries" $u_i$ in $U$ are distinct).
If we fix one such string $s$, the $2^{2n-1}$ pairs $(u, u \oplus s)_{u \in \{0, 1\}^{2n}}$ are determined, and the permutation 
defining $A$ at length $2n$ must be chosen so as to map $u_i$ to $a_i$ for all $u_i \in U$. The number of
such permutations does not depend on the fixed string $s$.
Thus, the numerator is equal to the probability over $A$ that $s_{2n, A}$ is in the 
set $\{u_j \oplus u_i~|~ u_i \in U \} \, \setminus \,
\{u \oplus v ~|~ u,v \in U \mbox{ and } u \neq v  \}$ times the
probability that (for fixed $s$) 
the permutation defining $A$ at length $2n$ is 
compatible with $f_A(u_i) = a_i$, 
$u_i \in U$ (a probability that as noted above is 
the same for each $s$). The first factor is at most
\[
\frac{||U||}{2^{2n}-1} \leq \frac{j-1}{2^{2n}-1}.
\]
Similarly, the denominator in equation~(\ref{eq3}) is equal to the probability that $s$ is a string of 
length $2n$ different from $0^{2n}$
and not in the set $\{u \oplus v~|~u,v \in U \mbox{ and } u \neq v \}$ times the  
probability that (for fixed $s$) the permutation defining $A$ at length $2n$ is compatible with $f_A(u_i) = a_i$, 
$u_i \in U$ (and thus the second factor of the denominator is the same as the second factor of the numerator). The first 
factor of the denominator is 
at least
\[
\frac{2^{2n} - 1 -||U|| (||U||-1) /2}{2^{2n} -1} \geq \frac{2^{2n} - 1 - (j-1)(j-2)/2}{2^{2n} -1}.
\]
Consequently, the fraction in equation~(\ref{eq3}) is bounded from above by
\[
\frac{j-1}{2^{2n} - 1 - \frac{(j-1)(j-2)}{2}}.
\]

Substituting in equation~(\ref{eqsplit}), we obtain that
\begin{multline*}
\prob(x_j \mbox{ collides with $x_1$ or $x_2$ or $\ldots$ 
or $x_{j-1}$} \mbox{ and no collision 
in $\{x_1, \ldots, x_{j-1}\}$}) \\
\leq \frac{j-1}{2^{2n} - 1 - \frac{(j-1)(j-2)}{2}}\sum 
\prob( (\forall i \in \{1, \ldots, j  \})[x_i = u_i]\mbox{ and } (\forall i \in \{1, \ldots, j -1 \}) [f_A(u_i) = a_i]) \\
\leq
\frac{j-1}{2^{2n} - 1 -  \frac{(j-1)(j-2)}{2}}.
\end{multline*}

Thus, returning to equation~(\ref{eq1}), we obtain that
\begin{equation*}
\begin{split}
\prob(k \leq 2^{n/4} \mbox { and collision in } \{x_1, \ldots, x_k\})
&  \leq 
\sum_{j=2}^{2^{n/4}}  \frac{j-1}{2^{2n} - 1 - \frac{(j-1)(j-2)}{2}} \\
& \leq \sum_{j=2}^{2^{n/4}} \frac{2^{n/4}}{2^{2n}-1 - (2^{n/2} - 1)} \\
& \leq \frac{2^{n/2}}{2^{2n} - 2^{n/2}} = \frac{1}{2^{3n/2}-1} \leq \frac{1}{2^{1.4n}},
\end{split}
\end{equation*}
which ends the proof of Fact~\ref{f:main-fact}.~\qed

We  mention that though it sometimes
happens in 
complexity theory that function results immediately
yield corresponding language results, 
it is not the case that our main result implies, at least in 
any obvious way, the corresponding language 
result.\footnote{Let
us be more explicit.
One might well wonder:

``It seems that your function result will
easily give the analogous language result.
Why?  Basically, by using the 
standard way we coerce function complexity into 
language complexity, i.e., via making a language that 
slices out bits or that prefix searches.
For example, using the first of these approaches,
take your hard function, call it $g$.  Now consider the function
$h$ defined as $h(\langle y,i \rangle) =$ the $i$'th bit of $g(y)$.  Since $g$
truth-table reduces to $h$, it follows that if $h$ has fast algorithms
then $g$ has fast algorithms (the relation depending on the length of
the query strings and the number of queries, but in fact in our case
these are such that one could make a good claim).  But you prove/claim that
$g$ does not have fast classical bounded-error algorithms, so neither
can $h$.  And certainly (this actually is the case) Brassard-H{\o}yer
easily still gives us that $h$ is quantum-easy to compute.''

However, this reasoning is not valid.  The above argument would
be fine if we were dealing with infinitely-often hardness.  However,
we are seeking to prove almost-everywhere hardness, and in fact the
bit-slices of an a.e.-hard function are not necessarily a.e.-hard.
To see this, consider any a.e.-hard function and prefix a 1 to 
all its
outputs.  This is still a.e.-hard but its bit-slices are
infinitely often trivial, namely, the first bit of each output is~1.
Of course, our hard function does not seem to 
have any such ``obvious'' or easy bits, but this is just an informal,
tempting hope rather than a valid proof.
} %

Another observation is that the proof of Theorem~\ref{t:our-main} actually shows the following
stronger result.

\begin{theorem}
\label{t:extension}
There is a constant $\epsilon > 0$ and a function oracle ${\cal O}$ relative to which there
is a problem $B$ computable in exact quantum polynomial time such that if $M$ is any bounded-error
classical Turing machine, then on all but a finite number of inputs $w$ on which the machine correctly
solves $B$, $M$ requires more than $2^{\epsilon |w|}$ steps.
\end{theorem}

In other words, even classical machines that are allowed to err infinitely
many times in their computation of the problem $B$ still need more
than $2^{\epsilon n}$ time on almost every input on which they are correct. The result follows
immediately from equation~\ref{e:prob1} and from the simulation of bounded-error machines by 
deterministic machines, both relativized with a random oracle, via the Bennett-Gill technique.

\medskip

\noindent{\bf Acknowledgments:}  We thank Andris Ambainis, Gilles Brassard,
Peter H{\o}yer, and Ronald de~Wolf for valuable literature pointers, comments,
and advice.

\singlespacing

\newcommand{\etalchar}[1]{$^{#1}$}

\end{document}